# On Improving Nuclear Fuel Imaging Using Position Sensitive Detectors

Erik Brücken[*,1], Peter Andersson[2], Mihaela Bezak[1], Peter Dendooven[1], Sofia Godø[4], Stefan Holm[2], Matti Kalliokoski[1], Aage Kalsæg[4], Gustav Pettersson[3], Anders Puranen[3], Vikram Rathore[2], Santeri Saariokari[1]

[1] Helsinki Institute of Physics, University of Helsinki, Finland
[2] Uppsala University, Department of Physics and Astronomy, Uppsala, Sweden
[3] SVAFO, Studsvik, Sweden
[4] IDEAS, Oslo, Norway

**ABSTRACT**

Spent nuclear fuel imaging before disposal is of utmost importance before long term disposal in dedicated storage facilities. Passive Gamma Emission Tomography (PGET) is an approved method by the International Atomic Energy Agency. The present detection system is based on small CZT detectors behind a tungsten-based collimator consisting of a linear array of slits. Small scale CZT crystals limit the detection efficiency of high energetic gamma rays from the fuel rods, mainly the 662 keV emissions from Cs-137. In our study based on full Monte-Carlo simulations as well as on experiments, we explore the capabilities of large pixelated CZT detectors to be used for PGET. We will discuss the theoretical advantages and practical challenges of the larger crystals. We demonstrate that the larger crystals, depending on their orientation, will increase the detection efficiency by a factor of 7 to 13. Due to the pixelated sensor signal readout we also explore the possibility to employ Compton imaging to improve the information on the location of origin of gamma rays.

In addition we explore the usefulness of commercial gamma-ray imagers for waste characterisation and decommissioning. In particular we report on the performance of the GeGI imager from PHDS Co and the H420 imager from H3D Inc in measuring nuclear waste drums at Svafo, Sweden.

## 1    INTRODUCTION

The interest in nuclear fuel imaging is rising, especially in Finland that has taken the leading role in developing a geological repository for the long term storage for spent nuclear fuel. One particularly useful device for nuclear safeguards for spent fuel assemblies (SFA) is the Passive Gamma Emission Tomography (PGET) device which has its development roots back in the 1980s. In 2017 it was approved by IAEA [1] for safeguard inspections. However, further development of the PGET device is ongoing. The current system is based on small Cadmium Zinc Telluride detectors (CZT). Those are placed precisely behind slits of a collimator made of tungsten. Two identical collimator/detector systems are positioned opposite each other on a rotating platform within a watertight enclosure, submerged underwater for scanning the fuel assemblies. The current PGET device is shown in Figure 1. This system is limited by the small CZT detectors with relatively low efficiency for full gamma-ray energy detection. This is especially crucial for imaging the centre positions of a SFA as the typical 662 keV gamma-rays form the centre are strongly attenuated.

Similar limitations concern the detection of the higher energetic gamma-rays of 1274 keV from Eu-154 which are less intense in SFAs.

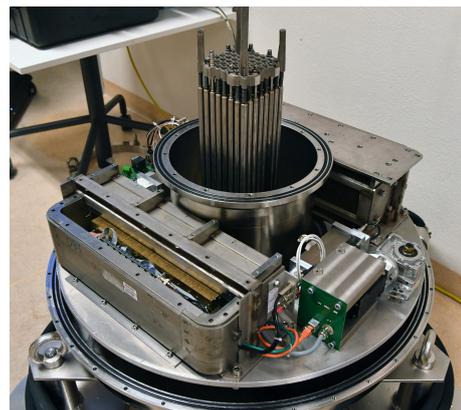

*Figure 1: Current PGET device in open position. Credit: Dean Calma/IAEA.*

The first part of this paper reports our activities for improving the detection efficiency of PGET by utilizing larger pixelized CZT detectors [2]. We explore also the possibility of applying Compton imaging for enhancing detection efficiency and for detecting the point of origin of the gamma emissions. We employed detailed Monte-Carlo simulations using the GEANT4 toolkit. For comparison we conducted tomographic scans of mock-up fuel rods at Uppsala University.

---
[**] Email: jens.brucken@helsinki.fi

The second part explores the capabilities for nuclear waste characterisation of two commercial gamma-ray imagers based on slightly different technologies.

## 2   EVALUATION OF LARGE CZT DETECTORS FOR PGET

As main objective we study whether larger position-sensitive CZT detectors are performing better than the small detectors in the current PGET system in the context of SFA imaging. Note that the large detectors will cover several slits of the collimator with slit widths of 1.5 mm and a pitch of 4 mm. In order to be able to determine which slit the gamma-ray passed through the detectors need to be position-sensitive.

### 2.1   Simulations

We performed GEANT4-based simulations to evaluate whether large detectors enhance the detection performance of PGET [3]. We implemented large CZT detectors following the GDS-100 detector design from IDEAS AS, Norway, and placed them behind the model of a tungsten collimator with identical specifications of the PGET device. Figure 2 shows the geometry implementation of the simulation (top left) with the detectors positioned behind the collimator and a fuel rod in front. Different detector configurations are shown on the bottom row with the large CZT pixel detectors in two different orientation on the left and in the middle. The present configuration with small CZTs is shown on the right side.

The fuel rod was set to emit gamma-rays at two different energies, 662 keV and 1274 keV to resemble Cs-137 and Eu-154 isotopes. We generated 200 million events. The simulation was optimised to avoid photons not interacting with our set-up. It has to be noted that the simulation at this stage is not a full detector simulation. Detector efficiencies and signal digitisation are not taken into account.

Results of the simulation are shown in Figure 3. The energy spectrum on the top shows the deposited energy of the 662 keV photons for the three different detector configurations. A similar histogram is shown on the bottom for the 1274 keV photons. One can clearly see the increase of efficiency using larger CZT detectors. All usual features are shown such as the photo peak, Compton edge as well as characteristic x-ray peaks and annihilation peak at 511 keV in case of the 1274 keV photons. In terms of the large pixelized crystals, orientation 1 shows a slightly higher efficiency.

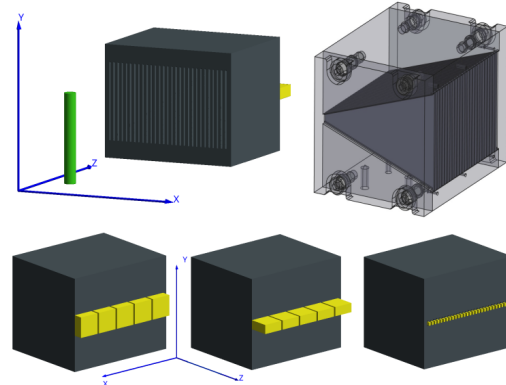

*Figure 2: Top left: Geometry implementation of the GEANT4 simulation showing the detectors behind the collimator and a fuel rod; top right: drawing of the collimator; bottom: different detector configurations behind the collimator (left: orientation 2, centre: orientation 1, right: current).*

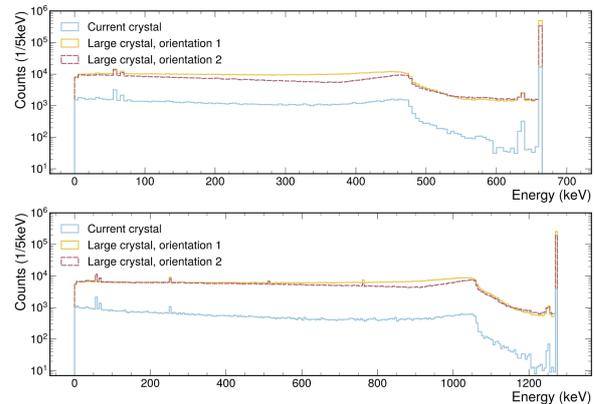

*Figure 3: Energy spectra of the Cs-137 662 keV (top) and the Eu-154 1274 keV photons (bottom) for the various detector configurations.*

Quantitatively the total 662 keV (1274 keV) gamma-ray interaction probability is about 11 (13) and 7 (10) times higher for the larger crystals in orientation 1 and 2 respectively compared to the smaller crystals. This is due to the larger detector thickness for the detector orientation 1 (22 mm) compared to case 2 (10 mm) in the direction of the incoming gamma-rays.

In terms of Compton imaging we looked at events that undergo Compton scattering followed by full photoelectric absorption of the scattered photon elsewhere in the detector. We found out that the fraction of these so-called golden events is up to 14% of the total number of interactions. We found no significant difference between the two orientations of the large crystal systems.

## 2.2 Tomographic Measurements

In collaboration with IDEAS we tested such a pixelized CZT detector system on a tomographic test bench at Uppsala University, Sweden, using mockup fuel rods. The position-sensitive detectors are made of 22x22x10 mm³ large CZT crystals with 11x11 anode pixels of 1.89 mm pitch and a planar cathode pad. A custom-made collimator of tungsten alloy (see drawing in Figure 2, top right) with slit and pitch values matching the PGET device was used. The two used mockup fuel rods contained bronze granulate enhanced with Cs-137 confined in a titanium tube, with an activity ratio of 0.548±0.018. The setup is shown in Figure 4. After initial calibration data acquisition runs we took two tomographic runs. The detector holding two crystals (operating one crystal at time) was moved laterally to two positions to cover the full scanning area. At each position an additional 2 mm movement was taken to improve the position resolution due to the 4 mm pitch of the collimator. The rotational stage carrying two fuel rods was rotated in 3° steps.

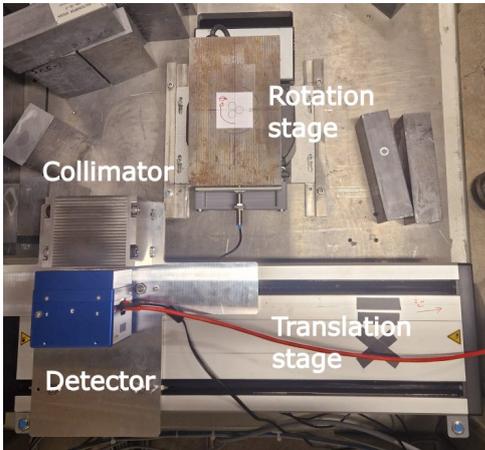

*Figure 4: Photograph of the tomographic test bench at Uppsala University.*

Results are shown in Figure 5. The image reconstruction is based on the filtered back projection algorithm using the Hann filter. Due to the different pixel pitch in comparison to the pitch of the collimator slits, slit efficiency differences were taken into account for the reconstruction. We find a good agreement between measurement and simulation also in terms of reconstructing the fuel rod activity ratio. During data taking we suffered from the relative high threshold of around 200 keV on the energy reconstruction that we had to apply due to high induced noise during data taking. This limited the capability of probing Compton imaging using the recorded data.

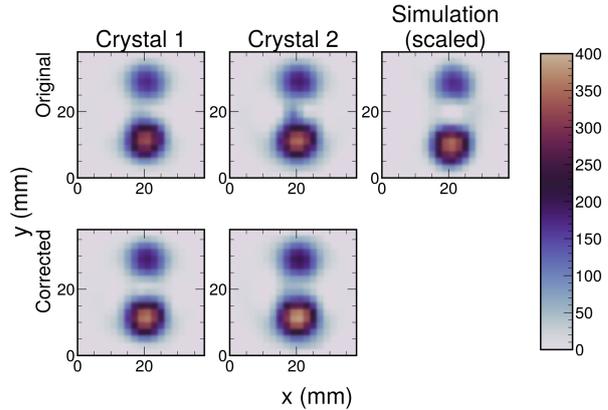

*Figure 5: Reconstructed images of the tomographic scans of the mock-up fuel rod assembly including comparison to simulation.*

## 3 WASTE CHARACTERISATION USING GAMMA-RAY IMAGERS

We performed measurements of nuclear waste drums at Svafo, Sweden using two commercially available gamma imagers from PHDS Co and H3D Inc [2].

### 3.1 Imagers

The **GeGI** imager from PHDS is a position sensitive semiconductor detector based on a Peltier-cooled high purity germanium disc with a diameter of 90 mm and a thickness of 11 mm. The double-sided strip sensor has 16 strips in orthogonal configuration, each with a pitch of 5 mm. It comes with an energy resolution of < 0.3% (FWHM) at 662 keV and a position resolution of 1.5 mm. The device operates in Compton imaging mode to localize identified radionuclides. Identification is based on recorded spectra using built-in radionuclide libraries. Identification difficulties arise from sum peaks. Visualisation of unassigned emissions is possible based on energy interval analysis.

The **H420** imager from H3D is also a position-sensitive semiconductor detector but based on 4 CZT crystals each with a size of 22x22x10 mm³. Each crystal has 11x11 anode pixels with 1 cathode pad. The detector comes with an energy resolution of < 1.1% (FWHM) at 662 keV and a position resolution of 0.5 mm. This device also operates in Compton imaging mode similar to the GeGI device. In addition with its coded aperture it offers localisation detection using lower energy photon emissions.

## 3.2 Measurements

We first performed measurements of nuclear waste drums in a shielded chamber at Svafo. One of the waste drums as shown as example in Figure 6 has a size of 280 l. Inside the drum are insert drums of 200 l and 100 l size arranged such to create gaps in between. The gap of about 5 cm between the inner drum and the 200 l drum is filled with concrete and the gap to the outer drum with air. The inner drum itself contained contaminated and active nuclear waste.

The detectors were positioned directly in front of the drum in a central position with a distance of 80±10 cm as shown in Figure 2.

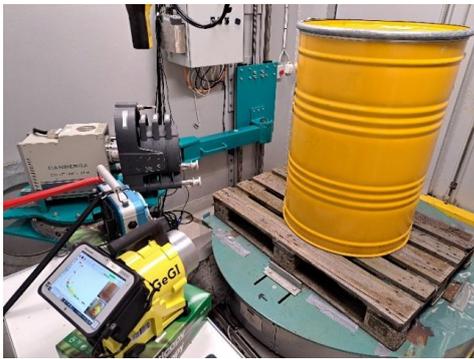

*Figure 6: Measurement of waste drum at Svafo.*

Both detectors come with digital cameras that allow to visualize the measured radiation overlayed on a photograph of the environment. In addition, a HPGe spectrometer was used to cross check the results of found radionuclides of the two gamma imagers. Drums were studied from up to 4 different angles, rotating the drums 90° each step.

## 3.3 Results

The two imagers performed relatively well in identifying and positioning radionuclides in the waste drums. Examples of identified radionuclides in drum 5 are shown in Table 1. Both imagers are capable of Compton imaging (Com) to localize the position of origin of radionuclides present. The H420 in addition has a coded aperture (CA) for localising radionuclides having lower energetic photon emissions. In Figure 7 the augmented visualisations of the GeGI and H420 imager are shown for the radionuclides Eu-152 and Cs-137 as example.

In general the results of both imagers were fairly comparable with a few exceptions. One exception was the detection of Co-60 where the two imagers had a discrepancy in positioning the source within the waste drum. We believe that it had to do with differences in detection capabilities of sources at low activity for the two imagers. In fact, the GeGI device was performing much better than the H420 in this respect. This is most likely due to the larger active area of the crystal (64 cm$^2$ vs. 19 cm$^2$) and the better energy resolution (0.3% vs. 1.1% at 662 keV).

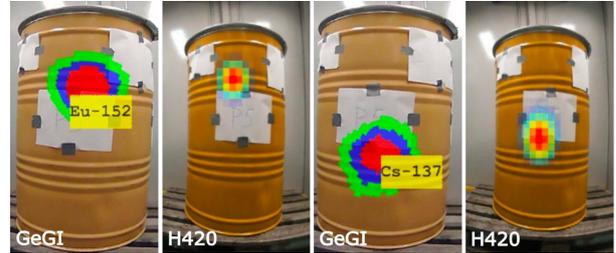

*Figure 7: Example measurement of drum 5 showing, among other detected isotopes, Eu-152 (left) and Cs-137 (right).*

*Table 1: Example results for drum 5 using the H420 and GeGI imagers.*

| Angle | Measurement Time [h] | Dead time | Dose rate [µSv/h] | Co-60 | Cs-137 | Eu-152 | Eu-154 |
|---|---|---|---|---|---|---|---|
| H420 imager | | | | | | | |
| 0° | 1 | 1.6% | 1.0 | Com | Com | Com/CA | Com |
| 180° | 16 | 1.7% | 1.6 | Com | Com | Com/CA | Com/CA |
| GeGI imager | | | | | | | |
| 0° | 1 | 0.4% | NA | Com | Com | Com | Com |
| 180° | 16 | 1.2% | NA | Com | Com | Com | Com |

## 4 CONCLUSIONS

We were able to show that position sensitive detectors can be used for tomographic scans of fuel rods and agreement with simulations is good. However, quantifying the improvement that justifies the replacement of the present PGET system is still to be performed. The rate capabilities of large CZT detectors as well as Compton imaging have to be further studied in this context.

Concerning the usage of commercial Compton imagers, we can conclude that such devices perform well within limitations. Low activity radionuclides are challenging and there are limitations due to the closed commercial software.

## ACKNOWLEDGEMENTS

This work was financially supported by NKS (Nordic nuclear safety research) under contract number AFT/NKS-R(24)136/4.